\documentclass[default]{sn-jnl}

\theoremstyle{thmstyleone}%
\theoremstyle{thmstyletwo}%

\theoremstyle{thmstylethree}%

\usepackage[utf8]{inputenc}
\usepackage{comment}
\usepackage{pifont}
\usepackage{pdflscape}
\usepackage{rotating}
\usepackage{booktabs}
\usepackage{soul}
\usepackage{longtable}
\usepackage{color, colortbl}
\definecolor{Gray}{gray}{0.7}
\usepackage{arydshln}

\begin{document}

\title[Article Title]{Automatic 3D Multi-modal Ultrasound Segmentation of Human Placenta using Fusion strategies and Deep Learning}


\author*[1]{\fnm{Sonit} \sur{Singh}}\email{sonit.singh@unsw.edu.au}

\author[2,3]{\fnm{Gordon} \sur{Stevenson}}\email{gordon.stevenson@unsw.edu.au}

\author[4]{\fnm{Brendan} \sur{Mein}}\email{brendan.mein@health.nsw.gov.au}

\author[2,3]{\fnm{Alec} \sur{Welsh}}\email{alec.welsh@unsw.edu.au}

\author[1]{\fnm{Arcot} \sur{Sowmya}}\email{a.sowmya@unsw.edu.au}

\affil*[1]{\orgdiv{School of Computer Science and Engineering}, \orgname{UNSW Sydney}, \orgaddress{\street{High St.}, \city{Kensington}, \postcode{2052}, \state{NSW}, \country{Australia}}}

\affil[2]{\orgdiv{School of Womens \& Childrens’ Health}, \orgname{UNSW Sydney}, \orgaddress{\street{High St.}, \city{Kensington}, \postcode{2052}, \state{NSW}, \country{Australia}}}

\affil[3]{\orgdiv{Department of Maternal-Fetal Medicine}, \orgname{Royal Hospital for Women}, \orgaddress{\street{Barkert St.}, \city{Randwick}, \postcode{2031}, \state{NSW}, \country{Australia}}}

\affil[4]{\orgdiv{Nepean Blue Mountains Local Health District}, \orgname{Nepean Hospital}, \orgaddress{\street{Derby St.}, \city{Kingswood}, \postcode{2747}, \state{NSW}, \country{Australia}}}

 \abstract{\textbf{Purpose:} Ultrasound is the most commonly used medical imaging modality for diagnosis and screening in clinical practice. Due to its safety profile, noninvasive nature and portability, ultrasound is the primary imaging modality for fetal assessment in pregnancy. Current ultrasound processing methods are either manual or semi-automatic and are therefore laborious, time-consuming and prone to errors, and automation would go a long way in addressing these challenges. Automated identification of placental changes at earlier gestation could facilitate potential therapies for conditions such as fetal growth restriction and pre-eclampsia that are currently detected only at late gestational age, potentially preventing perinatal morbidity and mortality. 
 \textbf{Methods:} We propose an automatic three-dimensional multi-modal (B-mode and power Doppler) ultrasound segmentation of the human placenta using deep learning combined with different fusion strategies. We collected data containing B-mode and power Doppler ultrasound scans for 400 studies.
 \textbf{Results:} We evaluated different fusion strategies and state-of-the-art image segmentation networks for placenta segmentation based on standard overlap- and boundary-based metrics. We found that multimodal information in the form of B-mode and power Doppler scans outperform any single modality. Furthermore, we found that B-mode and power Doppler input scans fused at the data level provide the best results with a mean Dice Similarity Coefficient (DSC) of 0.849.
 \textbf{Conclusion:} We conclude that the multi-modal approach of combining B-mode and power Doppler scans is effective in segmenting the placenta from 3D ultrasound scans in a fully automated manner and is robust to quality variation of the datasets.}


\maketitle

\section*{Introduction}\label{intro}
The \emph{placenta} has roles in fetal growth and development, oxygenation and nutrition, synthesising vital substances for pregnancy maintenance, including estrogen, progesterone, cytokines, and growth factors, and acting as a barrier against pathogens and drugs. Placental dysfunction is a leading cause of perinatal morbidity and mortality, including fetal growth restriction (FGR), pre-eclampsia, and stillbirth~\citep{Lean:2017:Placental_dysfunction}. The \emph{in vivo} assessment of placenta across gestation is critical to understand placental structure, function, and development and to identify strategies to optimise pregnancy outcome~\citep{Guttmacher:2015:the_human_placenta_project}.

The primary modality for placental evaluation is two-dimensional (2D) ultrasound (US), which is non-invasive, inexpensive and more easily acceptable and accessible than other imaging modalities such as X-ray or Magnetic Resonance Imaging (MRI). It may be used to characterize location, shape, and volume of the placenta along with its interface with the endometrium and myometrium. Three-dimensional Power Doppler (PD) ultrasound permits direct visualisation of multi-directional placental vascularity, allowing assessment of both the uteroplacental and fetoplacental circulations, providing dynamic assessment of blood flow for imaging of abnormalities of the placenta. 

\begin{figure}
    \centering
    \includegraphics[scale=0.7]{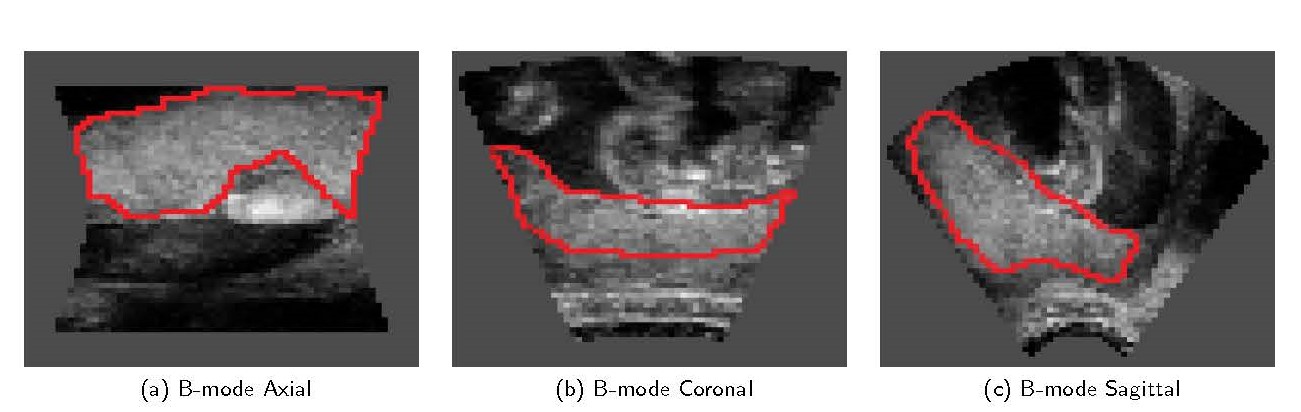}
    \caption{Visualisation of human placenta in three-dimensional ultrasound. Red contour shows placenta in axial, coronal, and sagittal views.}
    \label{fig:visualization_human_placenta}
\end{figure}

In three-dimensional (3D) ultrasound, a process called \emph{semantic segmentation} could be used to separate the placenta for qualitative and quantitative analysis. Placenta segmentation is challenging because of its geometry, position, and appearance as the shape and location of placentas vary greatly across subjects~\citep{Abulnaga:2019:placental_flattening} and fetal position can lead to shadowing artefacts. Determination of the placental boundary in relation to the uterine tissue is also challenging due to similar appearances~\citep{Looney:2017:automatic_3D_US_segmentation}, irregularity of boundary and changing size and shape with gestation, posing problems for segmentation~\citep{Han:2019:automatic_segmentation_of_human_placenta}. Figure~\ref{fig:visualization_human_placenta} shows 3D ultrasound providing placental visualisation in \emph{axial}, \emph{coronal}, and \emph{sagittal} views. It can be seen in the figure that the placenta is quite irregular and is texturally similar to other tissues of the uterine wall and other nearby organs.

Manual 3D-US placental segmentation is laborious, time-consuming, and has high inter/intra-observer variability, with the fundamental US signal having low image quality and being prone to speckle noise and being influenced by machine settings and operator interpretations.Automated placental segmentation and identification could revolutionise clinical practice~\citep{Diniz:2020:DL_strategies_for_US_in_pregnancy} by providing a tool to predict future pregnancy complications. Identification of placental changes at earlier gestation could facilitate potential therapies for conditions such as fetal growth restriction and pre-eclampsia that are currently detected only at late gestational age, potentially preventing perinatal morbidity and mortality. We propose that the use of multiple modalities (both grayscale and Doppler concurrently) will facilitate automated 3D placental US segmentation. Using different fusion strategies and deep learning, we applied our proposed methodology to a private dataset, to demonstrate its effectiveness. The main contributions of the research presented in this paper are:

\begin{enumerate}
    \item A novel fully automated placental segmentation methodology; 
    \item Use of different data fusion strategies (early fusion, multi-stage fusion, and late fusion) to combine the benefits of B-mode and power Doppler ultrasound scans;
    \item Demonstration of superiority of this proposed methodology over existing work on placenta segmentation.
\end{enumerate}

\section*{Materials and Methods}\label{MaM}
\begin{figure}
    \centering
    \includegraphics[scale=0.7]{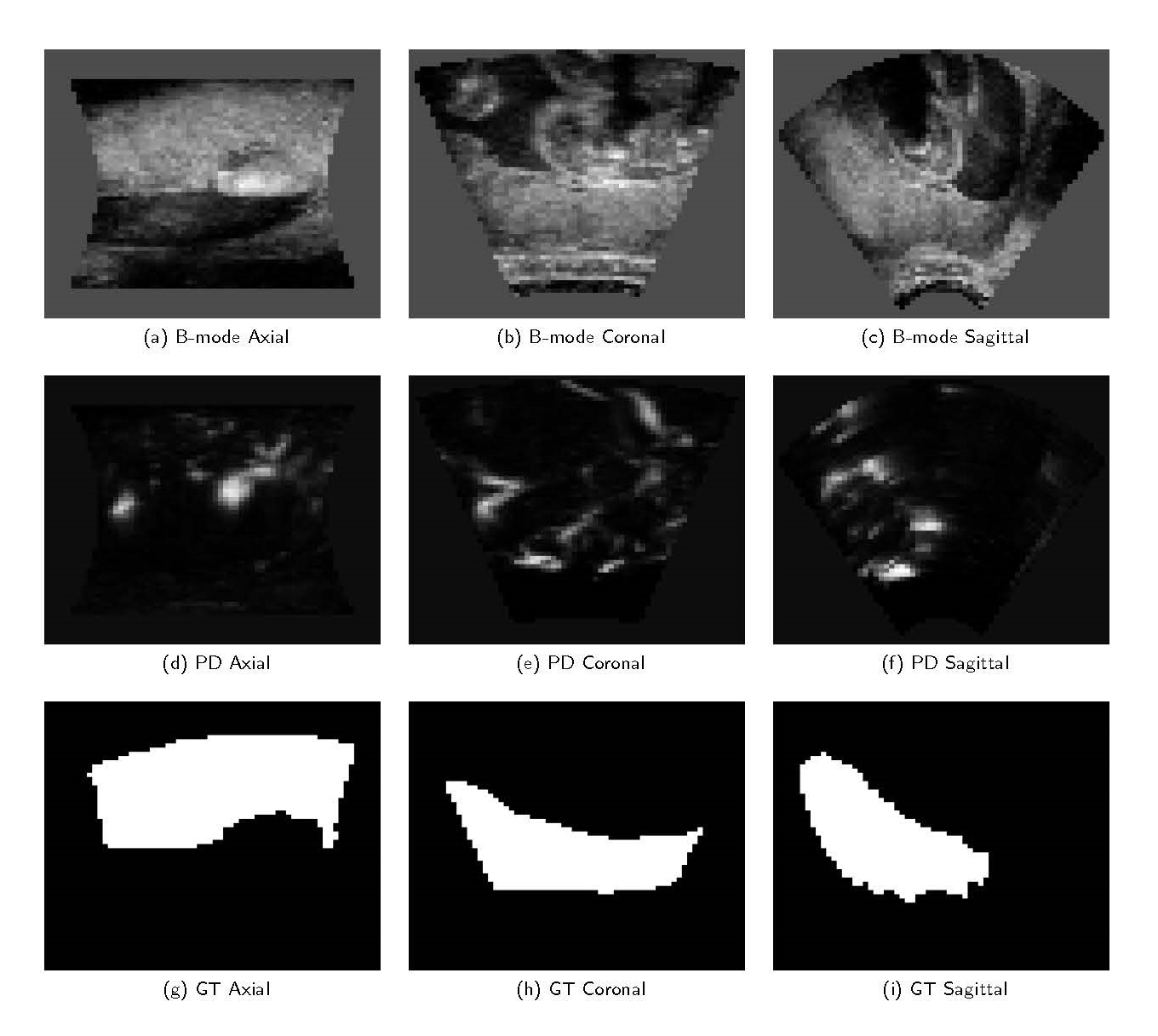}
    \caption{Sample dataset showing B-mode ultrasound (top row), Power Doppler (PD) ultrasound (middle row), and Ground Truth (GT) mask (bottom row) for the axial, coronal, and saggital views.}
    \label{fig:sample_dataset_1}
\end{figure}

\begin{figure}
    \centering
    \includegraphics[scale=0.7]{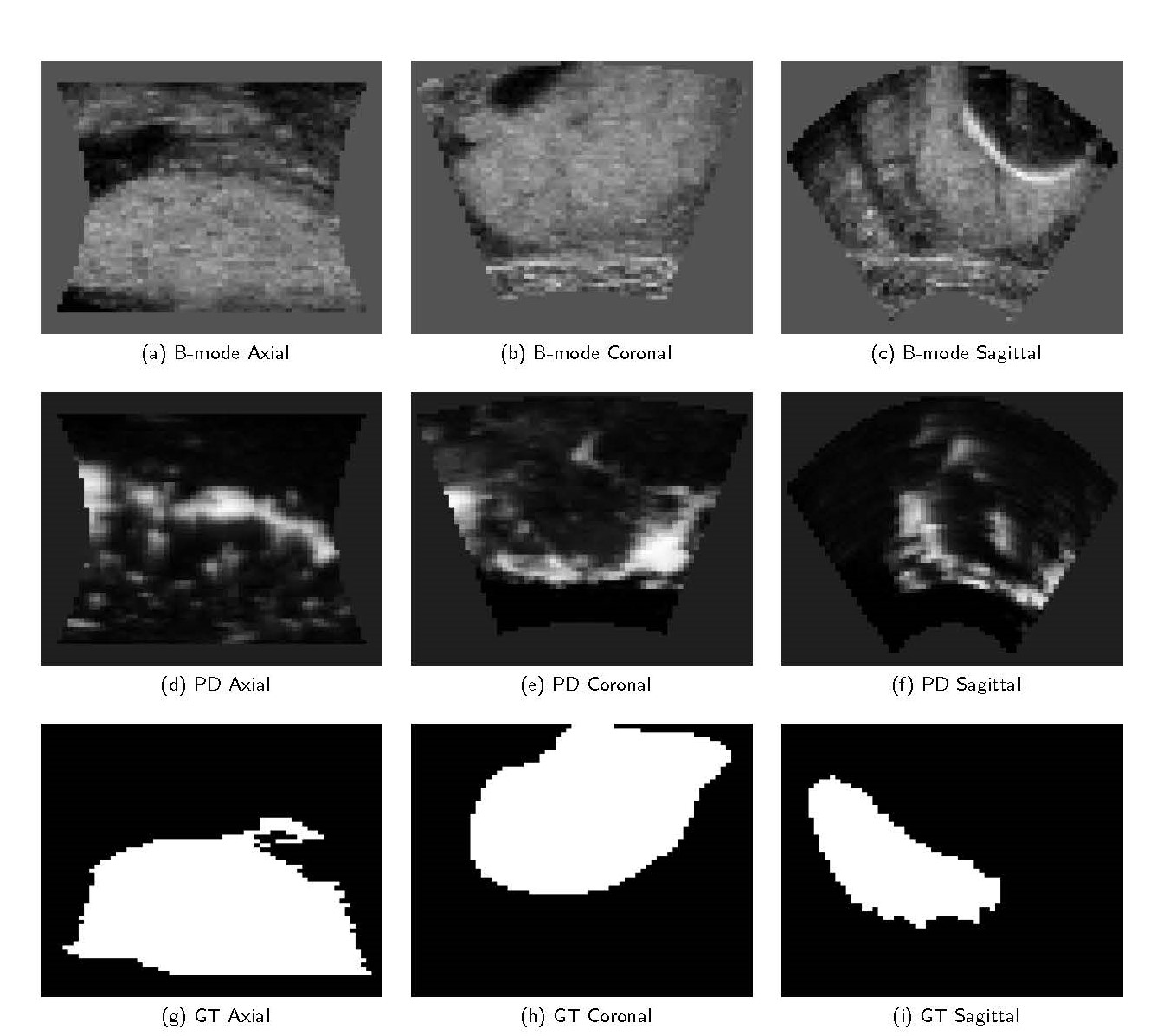}
    \caption{Sample dataset showing B-mode ultrasound (top row), Power Doppler (PD) ultrasound (middle row), and Ground Truth (GT) mask (bottom row) for the axial, coronal, and saggital views.}
    \label{fig:sample_dataset_2}
\end{figure}

This is a single-centre, retrospective, observational cohort study with local research ethics approval conducted with written informed consent. Anonymised data was collected from a study conducted at the Nepean Hospital under ethics approval from the Western NSW Local Health District (WNSWLHD) and the University of New South Wales (UNSW) Sydney. Participants were scanned by two experienced sonographers using a GE Voluson E8 ultrasound machine (GE Healthcare, Milwaukee, WI, USA) and a RAB4-8-D 3D/4D curved-array abdominal transducer (4-8.5 MHz). Data was available from 400 first-trimester 3D US pregnancy studies with B-mode and Power Doppler (PD). Raw 3D B-mode and PD US scans were exported with toroidal geometry (Kretz format; GE Healthcare, WI, USA) and converted to a 3D Cartesian volume for offline imaging analysis (Neuroimaging Informatics Technology Initiative (NIFTI; version 2) format. ITK-SNAP (version 3.6, University of Pennsylvania, PA USA) was used to visualise and manually segment the volumes. Three medical experts annotated by labelling voxels belonging to the placenta in 3D US B-mode volumes. We obtained manual segmentation and the best \emph{threshold} based segmentation mask using the following criteria: (1) using the segmentation mask as it was, in case there was only one segmentation mask; (2) computing the intersection of two volumes \emph{i.e.}, having voxel-based logical AND operation in case there are two annotated 3D ultrasound volumes; and (3) computing voxel-based majority voting in case there were three annotated 3D ultrasound volumes. Each B-mode scan had voxel spacing of $1.38114 \ mm \times 2.43458 \ mm \times 1.51483 \ mm$. For the segmentation framework, each 3D ultrasound volume (B-mode, PD, and segmentation mask) was converted to $64 \times 64 \times 64$ to have the same isotropic size. After this, each of these 3D ultrasound volumes was normalised by taking each voxel value converting to a floating-point number in the range between 0 to 1. Data was scaled for voxel values to have a zero mean and unit variance. Figure~\ref{fig:sample_dataset_1} and Figure~\ref{fig:sample_dataset_2} shows samples from the training set.

\begin{figure}
    \centering
    \includegraphics[scale=0.37]{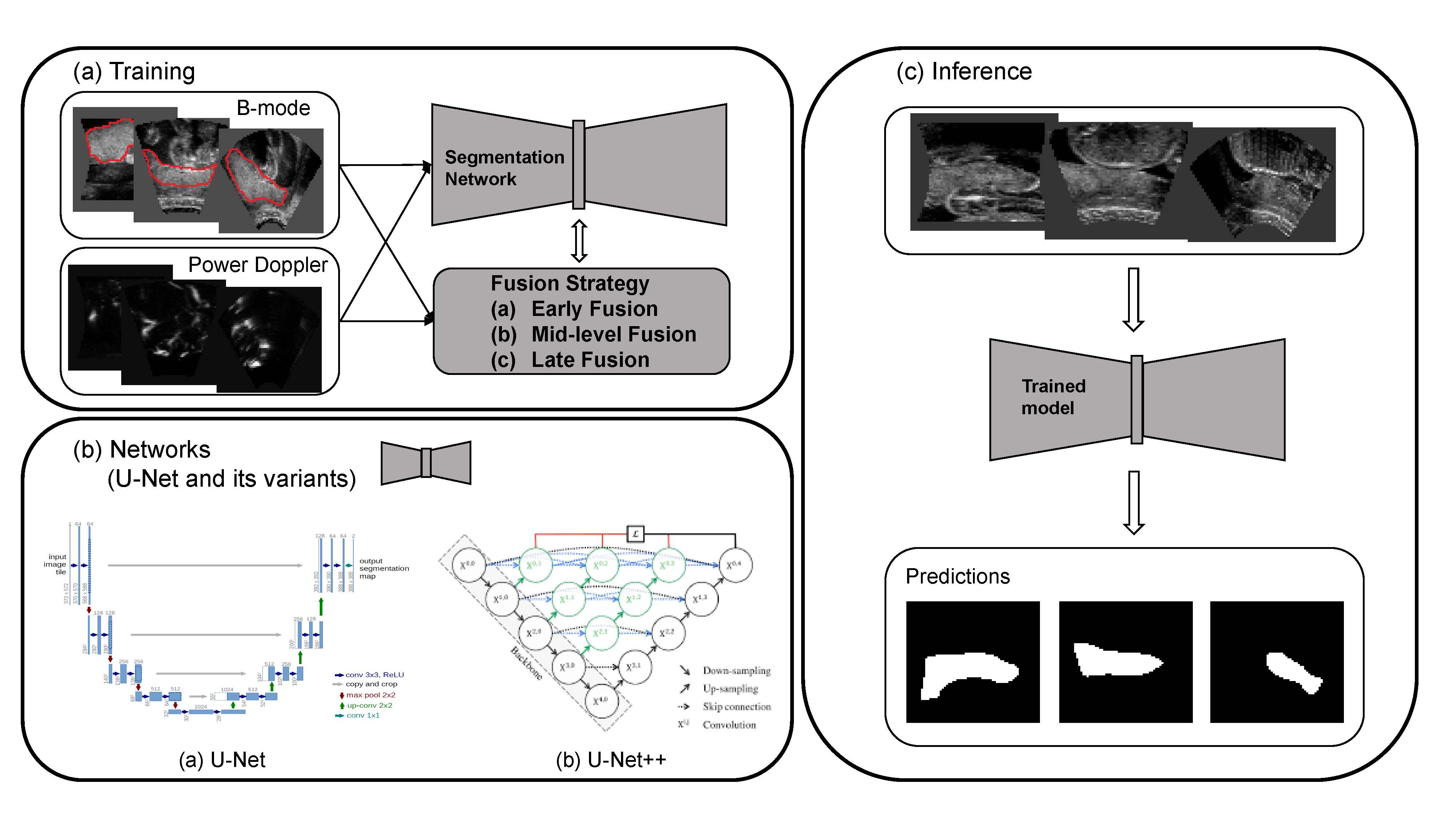}
    \caption{Block diagram of the proposed methodology. (a) Training data in the form of three-dimensional B-mode and Power Doppler scans and annotated masks for network training. Fusion strategies such as early fusion, multi-stage fusion, and late fusion are applied combining two modalities. (b) Different medical image segmentation networks such as U-Net, U-Net++, and their variants. (c). Model inference for the three-dimensional B-mode scan predicting mask for the axial, coronal, and the sagittal views.}
    \label{fig:block_diagram}
\end{figure}

Figure~\ref{fig:block_diagram} shows block diagram of our proposed methodology. We proposed combination of a 3D Convolutional Neural Network (CNN) architecture for segmentation with a series of fusion strategies to combine B-mode and power Doppler ultrasound images.  We adapted U-Net~\citep{Ronneberger:2015:U-Net}, a medical image segmentation network based on encoder-decoder framework. The encoder extracts image features through various convolutional layers whereas the decoder combines low-dimensional features to form high-dimensional features using the up-sampling process. Skip-connections, also called shortcut connections are also used that help to capture global and local information, beneficial for segmentation segmentation. In U-Net, each encoder block consists of two $3 \times 3$ convolution layers and each convolution layer is followed by a \emph{ReLU} activation and batch normalisation. This is followed by $2 \times 2$ \emph{max-pooling} which reduces the spatial dimensions of feature maps by half. Each decoder block consists of a $2 \times 2$ up-convolution layer. The output of up-convolution is concatenated with the skip connection feature map coming from encoder blocks. These skip connections provides global information from earlier layers of the encoder block as information is lost due to the depth of the network. After this, two $3 \times 3$ convolution were applied, where each convolution was followed by a ReLU activation function. The output of the last decoder passed through a $1 \times 1$ convolution with \emph{sigmoid} activation. Following the same encoder-decoder architecture, we also used U-Net++~\citep{Zhou:2018:UNet++} utilizing nested and dense skip connections. UNet++ can more effectively capture fine-grained details of the foreground objects when high-resolution feature maps from the encoder network are gradually enriched prior to fusion with the corresponding semantically rich feature maps from the decoder network. UNet++ is an extension of the original U-Net architecture where all convolutional layers along a skip pathway ($X_{i,j}$) use $k$ kernels of size $3 \times 3 \times 3$ where $k = 32 \times 2^i$. To enable deep supervision, a $1 \times 1$ convolutional layer followed by a \emph{sigmoid} activation was appended to each of the target nodes: $\{x_{0,j} \mid j \in \{1,2,3,4\}\}$ . As a result, U-Net generates four segmentation maps given an input image, which were further averaged to generate the final segmentation map.

\begin{figure}
    \centering
    \includegraphics[scale=0.4]{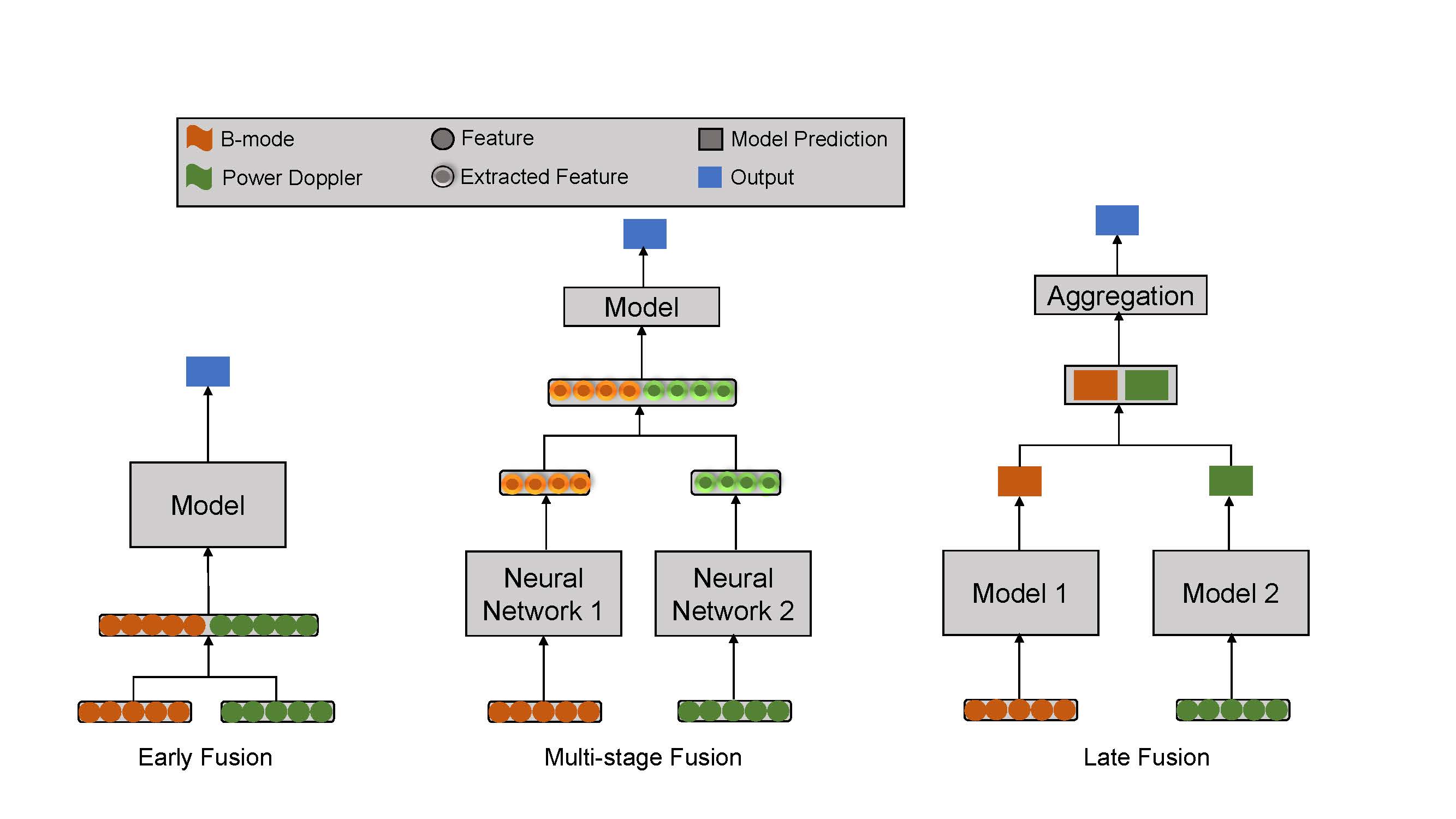}
    \caption{Model architecture for different fusion strategies. Early fusion (left) concatenates original features at the input level. Multi-stage or joint fusion (middle) concatenate extracted features. Late fusion (right) aggregates predictions at the decision level. Neural network in the diagram denotes image segmentation network.}
    \label{fig:model_architecture}
\end{figure}

Figure~\ref{fig:model_architecture} shows architecture for three fusion strategies we applied in this work, namely, \emph{early fusion}, \emph{multi-stage fusion}, and \emph{late fusion}. The aim of these fusion strategies was to effectively exploit complementary, redundant, and cooperative features of different modalities~\citep{Stahlschmidt:2022:Multimodal_DL}. In \emph{early fusion}, the original input data from two modalities are concatenated, and the resulting vector is treated like uni-modal input. The early fusion strategy can learn cross-modal relationships from low-level features. In \emph{intermediate fusion}, marginal representations in the form of feature vectors are learned and fused instead of the original multimodal data. In \emph{late fusion}, instead of combining the original data or learned features, decisions of separate models are combined into a final decision. The simplest approach to aggregating decisions from separate models is to take the average of the individual outputs. This could be the averaging the probabilities from softmax functions for each class. Data fusion techniques have been applied on multidimensional medical and biomedical data in detecting different diseases as multiple modalities can contribute more sufficiently than a single modality~\citep{Azam:2022:Review_data_fusion}. A data fusion can decrease prediction errors and increase the reliability of the interpretation. 

We implemented the proposed methodology in Python using Keras library with a Ubuntu workstation with Tensorflow backend with 16 cores of 2.80 GHz having Nvidia RTX 2080Ti GPU (12 GB memory) and CUDA 11.1 installed on it. For the segmentation networks, we tried increasing (16, 32, 64, 128, 256) feature maps per layer. All images including B-mode scans, PD scans, and ground-truth masks were re-sampled to $64 \times 64 \times 64$ isotropic size. To increase dataset size and add diversity in data during model training, we applied affine transformations (translation range of 10 pixels, rotation range of 10-degrees, scaling factor of 10, and shearing range of 15 voxels) to augment data during network training. The initial learning rate was $10^{-4}$, reduced by a factor of 0.1 after every 10 epochs. All networks were optimised using the ADAM optimiser and trained until convergence. Models were trained for 80 epochs. The network hyper-parameters were determined based on the best performance on the validation set. We performed a 5-fold cross-validation to report results with each fold dividing studies into training (60\%/240 studies), validation (20\%/80 studies), and test set (20\%/80 studies) sets. Within each set, we randomly shuffled images, keeping B-mode image, PD image, and the segmentation mask for each study in the same directory. During the data split process, we ensured that images from individual patients were not distributed across training, validation, and testing set, preventing any data leakage. Table~\ref{tab:percentage_dissimilarity} shows the percentage dissimilarity between different folds of the dataset, ensuring sufficient diversity. For each fold, we compared pair-wise data similarity between different folds for training, validation, and test set.

\begin{table}[]
    \centering
    \caption{Percentage dissimilarity between different folds of dataset.}
    \label{tab:percentage_dissimilarity}
     \begin{tabular}{|c|c|c|c|c|c|}
    \hline
      \textbf{Fold} & \multicolumn{5}{c}{Percentage dissimilarity (training, validation, test)} \\ \hline
         & \textbf{1} & \textbf{2} & \textbf{3} & \textbf{4} & \textbf{5} \\
    \hline
    \textbf{1} & (0,0,0) & (40.8,80,65) & (40.4,87.5,81.3) & (46.3,87.5,75) & (43.3,70,96.3) \\
    \textbf{2} & (40.8,80,65) & (0,0,0) & (49.6,91.3,100) & (43.3,90,82.5) & (32.1,82.5,91.3) \\
    \textbf{3} & (40.4,87.5,81.3) & (49.6,91.3,100) & (0,0,0) & (42.9,87.5,76.3) & (45.8,78.8,87.5) \\
    \textbf{4} & (46.3,87.5,75) & (43.3,90,82.5) & (42.9,87.5,76.3) & (0,0,0) & (55.8,96.3,91.3) \\
    \textbf{5} & (43.3,70,96.3) & (32.1,8.5,91.3) & (45.8,78.8,87.5) & (55.8,96.3,91.3) & (0,0,0) \\
    \hline
    \end{tabular} 
\end{table}

For segmentation performance, we used two overlap measures (Dice Similarity Coefficient (DSC) and Jaccard Similarity Index (Jaccard Index) and two surface measures (Hausdorff Distance (HD) and Mean Surface Distance (MSD)). Overlap-based metrics measure the overlap between the reference annotation (mask) and the prediction of the algorithm. Distance-based metrics measure the distance between object boundaries. The \emph{DSC} is a commonly applied index measuring the similarity between two sets of data. The \emph{Jaccard Index}, also known as the Intersection-over-Union (IoU) measures the number of pixels common between the ground-truth mask and the predicted mask divided by the total number of pixels present across both masks. The \emph{Hausdorff Distance} (HD) metric calculates the maximum of all shortest distances for all points from one object boundary, with the HD95 calculating the 95\% percentile to remove excess influence of outliers. The HD95 metric has limitations if one boundary is much larger than the other one. this may be overcome by the \emph{Mean Surface Distance} (MSD) score which treats both structures equally by computing the average distance from structure A to structure B and the average distance from structure B to structure A and averaging both. 

\section*{Results}\label{Results}
\begin{figure}
    \centering
    \includegraphics[]{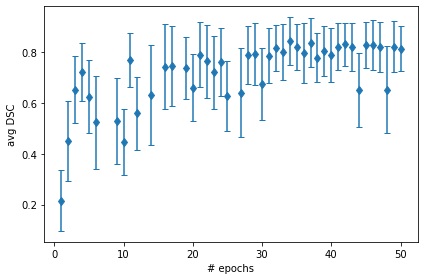}
    \caption{Segmentation results (\emph{i.e.}, average DSC) as a function of number of epochs needed for U-Net convergence when trained and validated on one fold of the dataset (\# training=240, \#validation=80, and \#test=80).}
    \label{fig:segmentation_results_epochs}
\end{figure}

Results for placenta segmentation using baseline U-Net are shown in Table~\ref{tab:segmentation_results_UNet}, reported by averaging metric values over all studies in the test set. Table~\ref{tab:segmentation_results_UNetpp} shows results for placenta segmentation using the U-Net++ model. These tables show that more robust networks such as U-Net++ provides better results by giving higher DSC score, higher Jaccard Index score, lower HD (in mm) score, and lower MSD (in mm) score. Table~\ref{tab:segmentation_results_with_augmentation} shows the effectiveness of applying data augmentation during network training. Both U-Net and U-Net++ networks showed better results with data augmentation compared to without any data augmentation, highlighting benefits of data augmentation by increasing dataset size and diversity during network training. Results for the three fusion strategies are shown in Table~\ref{tab:segmentation_results_fusion_strategies}, with \emph{early fusion} giving best results compared to \emph{intermediate} and \emph{late} fusion strategies. To show that networks implemented in this study were converging, we tracked average DSC on the performance of U-Net on the validation set, finding that the average DSC on the validation set gradually increased with increase in the number of training epochs, finally reaching a saturation point after 50 epochs as shown in Figure~\ref{fig:segmentation_results_epochs}. This guided us to run various networks in this study for 80 epochs to be sufficient for network convergence. During the network training, we saved the best model based on the validation set during network training, which was then used for network inference and to report results on the unseen test set.

\begin{table}[]
    \centering
    \caption{Segmentation results comparing U-Net model performance for five folds of the final dataset with each fold having \#train=240, \#validation=80, and \#test=80 3D ultrasound volumes. Results are averaged values over all studies in the test set with $\pm$ standard deviation of metric for that test set.}
    \label{tab:segmentation_results_UNet}
    \begin{tabular}{c|cccc}
    \hline
     \textbf{Fold\# (Dataset)} & \textbf{DSC} & \textbf{Jaccard Index} & \textbf{HD (mm)}  & \textbf{MSD (mm)} \\
     \hline
      Fold 1 & 0.823 $\pm$ 0.101 & 0.708 $\pm$ 0.102 & 8.645 $\pm$ 6.322 & 1.501 $\pm$ 0.454 \\
      Fold 2 & 0.825 $\pm$ 0.058 & 0.706 $\pm$ 0.076 & 7.920 $\pm$ 4.665 & 1.595 $\pm$ 0.631 \\
      Fold 3 & 0.823 $\pm$ 0.064 & 0.704 $\pm$ 0.082 & 10.500 $\pm$ 6.111 & 1.664 $\pm$ 0.887 \\
      Fold 4 & 0.814 $\pm$ 0.075  & 0.692 $\pm$ 0.093 & 7.978 $\pm$ 4.839 & 1.722 $\pm$ 0.912 \\
      Fold 5 & 0.821 $\pm$ 0.045 & 0.698 $\pm$ 0.062 & 8.262 $\pm$ 4.420 & 1.572 $\pm$ 0.408 \\
    \hline
    \end{tabular}
\end{table}

\begin{table}[]
    \centering
    \caption{Segmentation results comparing U-Net++ model performance for five folds of the final dataset with each fold having \#train=240, \#validation=80, and \#test=80 3D ultrasound volumes. Results are averaged values over all studies in the test set with $\pm$ standard deviation of metric for that test set.}
    \label{tab:segmentation_results_UNetpp}
    \begin{tabular}{c|cccc}
    \hline
     \textbf{Fold\# (Dataset)} & \textbf{DSC} & \textbf{Jaccard Index} & \textbf{HD (mm)}  & \textbf{MSD (mm)} \\
     \hline
      Fold 1 & 0.828 $\pm$ 0.076 & 0.706 $\pm$ 0.067 & 4.898 $\pm$ 3.156 & 1.196 $\pm$ 0.752 \\
      Fold 2 & 0.819 $\pm$ 0.042 & 0.694 $\pm$ 0.113 & 7.348 $\pm$ 5.245 & 2.039 $\pm$ 0.574 \\
      Fold 3 & 0.824 $\pm$ 0.056 & 0.700 $\pm$ 0.018 & 7.615 $\pm$ 4.758 & 1.502 $\pm$ 0.626 \\
      Fold 4 & 0.833 $\pm$ 0.107  & 0.715 $\pm$ 0.045 & 4.690 $\pm$ 3.167 & 1.340 $\pm$ 0.285 \\
      Fold 5 & 0.840 $\pm$ 0.072 & 0.725 $\pm$ 0.057 & 4.123 $\pm$ 3.032 & 1.177 $\pm$ 0.377 \\
    \hline
    \end{tabular}
\end{table}

\begin{table}[]
    \centering
    \caption{Segmentation results with and without data augmentation. Results are averaged values over all studies in the test set with $\pm$ standard deviation of metric for that test set.}
    \label{tab:segmentation_results_with_augmentation}
    \begin{tabular}{p{3cm}p{1.8cm}p{1.8cm}p{1.8cm}p{1.8cm}}
    \hline
     \textbf{Method} & \textbf{DSC} & \textbf{Jaccard Index} & \textbf{HD (mm)}  & \textbf{MSD (mm)} \\
     \hline
      U-Net, without data augmentation & 0.824 $\pm$ 0.074 & 0.700 $\pm$ 0.080 & 7.615 $\pm$ 4.674 & 1.502 $\pm$ 0.571 \\  \hdashline
      U-Net, with data augmentation & 0.833 $\pm$ 0.043 & 0.714 $\pm$ 0.037 & 4.690 $\pm$ 3.247 & 1.340 $\pm$ 0.164 \\ \hdashline
      U-Net++, without data augmentation & 0.839 $\pm$ 0.068 & 0.722 $\pm$ 0.056 & 7.141 $\pm$ 4.635 & 1.279 $\pm$ 0.386 \\ \hdashline
      U-Net++, with data augmentation & 0.847 $\pm$ 0.036 & 0.725 $\pm$ 0.079 & 4.123 $\pm$ 3.752 & 1.177 $\pm$ 0.296 \\
    \hline
    \end{tabular}
\end{table}

\begin{table}[]
    \centering
    \caption{Segmentation results applying applying early fusion, intermediate fusion, and late fusion for the two modalities, namely, B-mode and Power Doppler 3D ultrasound volumes. Results are averaged values over all studies in the test set.}
    \label{tab:segmentation_results_fusion_strategies}
    \begin{tabular}{p{5.5cm}p{1cm}p{1cm}p{1cm}p{1cm}}
     \hline
     \textbf{Method} & \textbf{DSC} & \textbf{Jaccard Index} & \textbf{HD (mm)}  & \textbf{MSD (mm)} \\
      \hline
      Early fusion (U-Net, without data augmentation) & 0.831 & 0.711 & 8.944 & 1.078 \\
      Intermediate fusion (U-Net, without data augmentation) & 0.825 & 0.702 & 5.196 & 1.484 \\
      Late fusion (U-Net, without data augmentation) & 0.818 & 0.693 & 9.110 & 1.582 \\ \hdashline
      Early fusion (U-Net++, without data augmentation) & 0.847 & 0.725 & 4.472 & 1.137 \\
      Intermediate fusion (U-Net++, without data augmentation) & 0.831 & 0.710 & 5.830 & 1.442 \\
      Late fusion (U-Net++, without data augmentation) & 0.826 & 0.704 & 16.643 & 1.229 \\ \hdashline
      Early fusion (U-Net, with data augmentation) & 0.838 & 0.722 &  4.898 & 1.144 \\
      Intermediate fusion (U-Net, with data augmentation) & 0.829 & 0.708 & 7.071 & 1.818 \\
      Late fusion (U-Net, with data augmentation) & 0.822 & 0.698 & 7.549 & 2.229 \\ \hdashline
      Early fusion (U-Net++, with data augmentation) & 0.849 & 0.738 & 4.051 & 1.007 \\
      Intermediate fusion (U-Net++, with data augmentation) & 0.839 & 0.722 & 7.141 & 1.279 \\
      Late fusion (U-Net++, with data augmentation) & 0.835 & 0.717 & 10.295 & 1.025 \\ 
     \hline
    \end{tabular}
\end{table}

\begin{figure}
    \centering
    \includegraphics[scale=0.7]{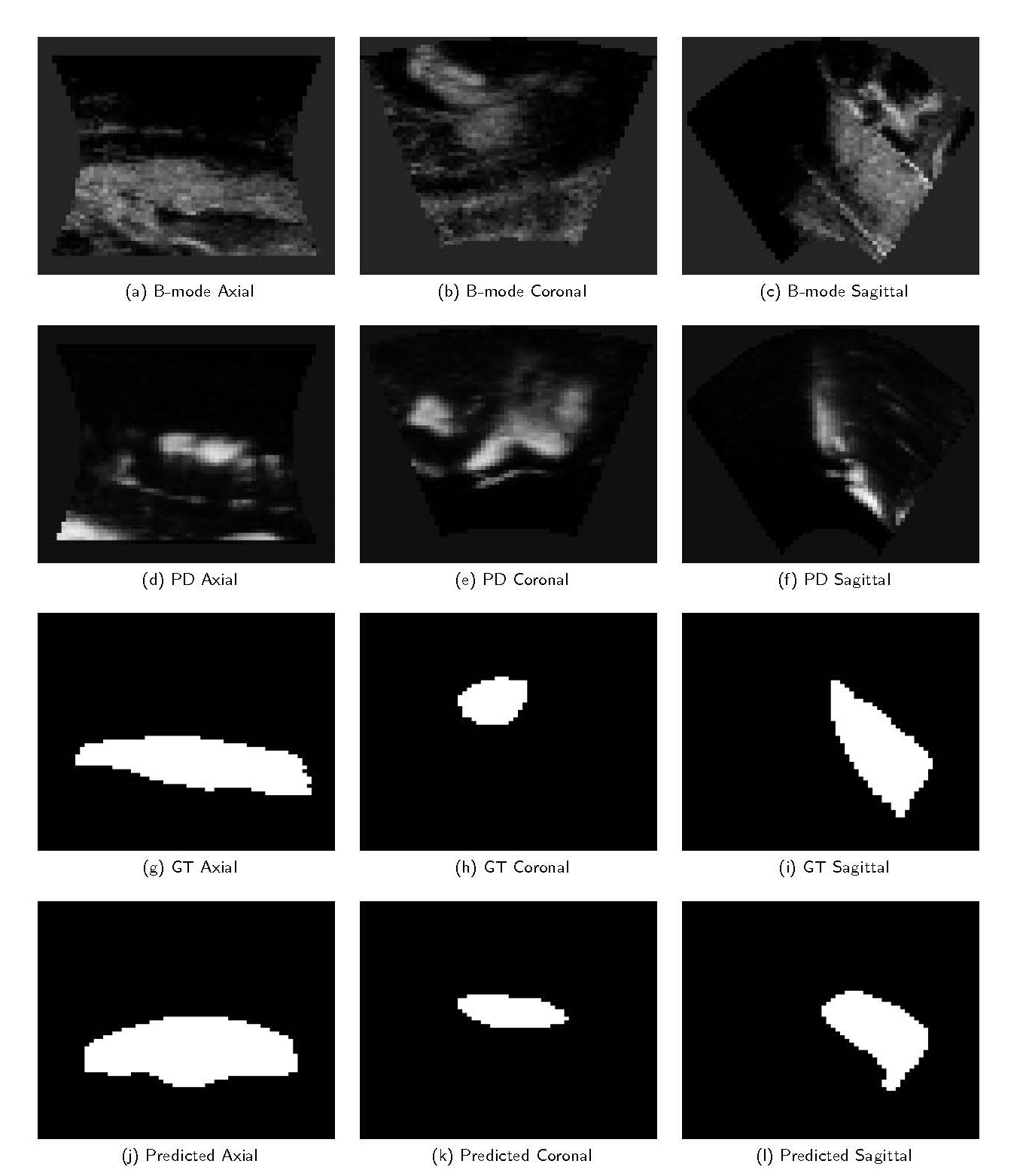}
    \caption{Qualitative results (Average case having DSC of 0.7135, Jaccard Index of 0.5546, HD95 of 9.4339, and MSD of 304.4385) showing Ground Truth (GT) mask (top row) the model predictions (bottom row) for the axial, coronal, and sagittal views.}
    \label{fig:qualitative_results_1}
\end{figure}

\begin{figure}
    \centering
    \includegraphics[scale=0.7]{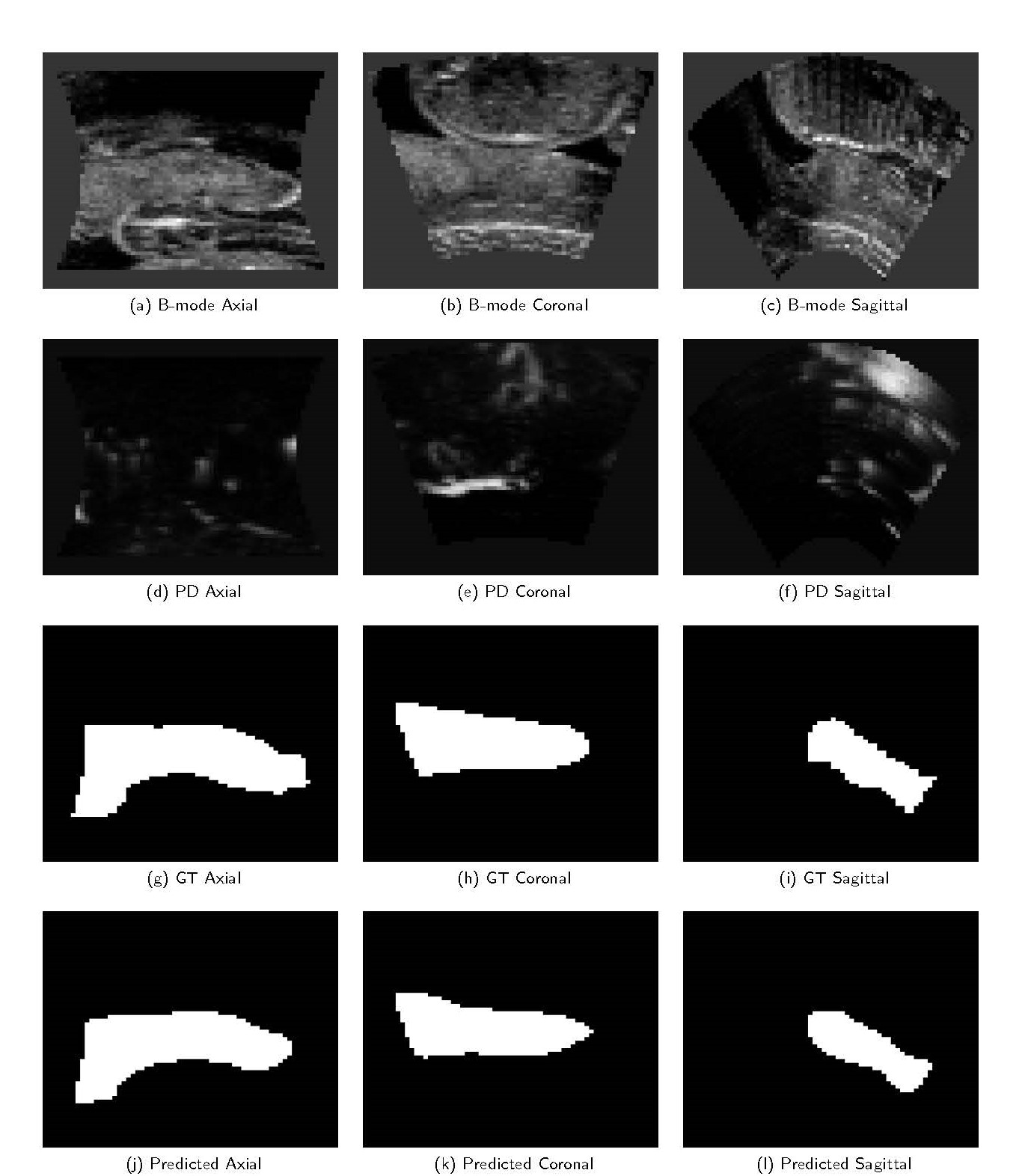}
    \caption{Qualitative results (Good case having DSC of 0.9039, Jaccard Index of 0.8247, HD95 of 3, and MSD of 0.7189) showing Ground Truth (GT) mask (top row) the model predictions (bottom row) for the axial, coronal, and sagittal views.}
    \label{fig:qualitative_results_2}
\end{figure}

Qualitative evaluation showed how the proposed methodology functioned for placental segmentation. As shown in Figure~\ref{fig:qualitative_results_1} and Figure~\ref{fig:qualitative_results_2} where the network had given an average DSC and high DSC, respectively, the model performed well in correctly identifying the centroid of the placenta region but struggled to correctly delineate the placental boundary having DSC of 0.7135, Jaccard Index of 0.5546, HD95 of 9.4339, and MSD of 304.4385. After detailed review with expert clinician feedback, we found that uterine regions surrounding the placenta have a similar texture, making it quite challenging for the model to correctly identify edges. This results in the misclassification of some non-placenta voxels as placenta voxels or vice-versa. We also found that placenta shows high variability in terms of its position, orientation, thickness, shape, and appearance. A study by \citet{Zimmer:2023:placenta_segmentation_US} showed that placenta segmentation in the \emph{posterior} orientation is challenging compared to placenta segmentation in the \emph{anterior} orientation. Based on our analysis, we found that one of the common sources of false positives is the thin region which separates placental uterine boundary. Because of a very fine region and having small differences in echo-textural properties between placenta and uterus tissues, many false positives occur in this this region. On the other hand, Figure 8 demonstrates a random sample where the model performs well for placenta segmentation, having DSC of 0.9039, Jaccard Index of 0.8247, HD95 of 3, and MSD of 0.7189. In this case, despite placental irregularity the model still performed well for placental segmentation. Other studies have also highlighted the issue that the placenta and uterine tissues are poorly differentiated in early pregnancy~\citep{Oguz:2016:fully_automated_placenta_segmentation}. Apart from this, the lack of contrast makes it difficult to accurately detect the boundary because of presence of speckle noise during ultrasound acquisition. For temporal processing, uterine contractions can also affect the shape of the placenta.

\section*{Discussion}\label{Discuss}
Automatic placental segmentation is of crucial for monitoring feto-maternal. It can facilitate early diagnosis of \emph{in utero} abnormalities and prevent fetal morbidity and mortality. The proposed methodology in this study utilised complementary information from both the B-mode and the power Doppler 3D ultrasound scans for placenta segmentation, the latter providing information for vascular analysis. 

Results of our study demonstrated that B-mode and power Doppler scans provide complementary information, resulting in higher placental segmentation accuracy. \emph{Early fusion} provided superior results compared to \emph{intermediate fusion} or \emph{late fusion}. We believe that this is because both modalities (B-mode and power Doppler) provided maximum information at the data level, whereas information loss occurs at intermediate or decision level. 

We compared results of the proposed methodology with existing work for placental segmentation based on the related work. Whilst direct comparison of results was not possible because of not having same dataset and network settings, Table~\ref{tab:comparison_SOTA} provides an overview of where the proposed work stands compared to the existing work. The DSC score is reported on the test set as given in these studies. Although the interactive random-walker method proposed by \citet{Stevenson:2015:3D_US_segmentation_of_placenta} gave high DSC of 0.87 $\pm$ 0.13, it is a semi-automated method requiring extensive user input in 3D and is sensitive to subtle variations in its initialisation. The work by \citet{Looney:2017:automatic_3D_US_segmentation,Looney:2018:fully_automated_real_time_3D_US,Looney:2021:Fully_automated_3D_US_segmentation} has subsequently improved placenta segmentation results from having mean DSC of 0.73 to 0.85 $\pm$ 0.05, though these studies have limitations as results are reported comparing predictions with the results from the semi-automatic random-walker method. As the ground-truth annotations are predictions made by the random-walker method, the ground-truth masks are noisy and do not denote expert manual segmentations. Furthermore, in \citet{Looney:2021:Fully_automated_3D_US_segmentation}, post-processing in terms of morphological operations is applied to smoothen segmented boundary and to get rid of false positives or false negatives.  The \emph{multi-atlas label fusion} method proposed by \citet{Oguz:2018:DL_with_multi-atlas_fusion} achieved a DSC of 0.83 but that dataset was limited to anterior placentas only, whereas our dataset contains a mix of both anterior and posterior placentas. The results reported in \citet{Hu:2019:automated_placenta_segmentation} gave quite high DSC of 0.92 $\pm$ 0.04, but their dataset was 2D images compared to our study where 3D images are taken. Authors of this study proposed a shadow detection layer which helps to overcome ultrasound artefacts such as speckle noise, in turn resulting in high DSC score. We believe that such contribution of removing artefacts is significant and can be applied to any placenta segmentation. Our proposed methodology using the fusion strategy achieved higher DSC score compared to when only using single modality. Although, the results of this study are slightly lower compared to semi-automatic methods where medical experts provided the seed points for the algorithm to converge, the current study is unique in terms of applying different fusion strategies and showing the effectiveness of using power Doppler along with B-mode ultrasound scans. This is an important contribution given we can not only have better placental segmentation but also can analyse vascularity of the placenta, providing greater insights about placental health. 

\begin{table}[]
    \centering
    \caption{Comparing proposed work with previous work on placenta segmentation. The results on test set in terms of Dice Similarity Coefficient (DSC) are directly reported from the papers.}
    \label{tab:comparison_SOTA}
    \begin{tabular}{p{2.5cm}p{2cm}p{2cm}p{2cm}p{2.5cm}}
    \hline
     \textbf{Study} & \textbf{Modality} & \textbf{Number of studies} & \textbf{Method} & \textbf{DSC score}  \\
     \hline
     \citet{Stevenson:2015:3D_US_segmentation_of_placenta} & 3D B-mode & 88 & Random walker; Semi-automatic & 0.87 $\pm$ 0.13 \\ 
     \citet{Oguz:2016:fully_automated_placenta_segmentation} & 3D B-mode & 14 & Multi-atlas label fusion & 0.83 $\pm$ 0.005 \\ 
     \citet{Yang:2019:Towards_automated_semantic_segmentation} & 3D B-mode & 104 & 3D CNN+RNN & 0.638 \\ 
     \citet{Hu:2019:automated_placenta_segmentation} & 2D B-mode & 1364 images (247 subjects) & U-Net and shadow layer & 0.92 $\pm$ 0.04 \\ 
     \citet{Oguz:2020:Minimally_interactive_placenta_segmentation} & 3D B-mode & 47 & Multi-atlas label fusion; Semi-automatic & 0.8238 $\pm$ 0.063 \\ 
     \citet{Looney:2018:fully_automated_real_time_3D_US} & 3D B-mode & 2,393 & OxNNet & 0.81 $\pm$ 0.15 \\ 
     \citet{Torrents-Barrena:2019:Placenta_segmentation} & 3D B-mode & 61 & 3D Conditional GAN (cGAN) & 0.75 $\pm$ 0.12 \\ 
     \citet{Looney:2021:Fully_automated_3D_US_segmentation} & 3D B-mode & 2,093 & 3D CNN + Postprocessing & 0.85 $\pm$ 0.05 \\ 
     \citet{Zimmer:2023:placenta_segmentation_US} & 3D B-mode & 292 & 3D Multi-task CNN & 0.87 $\pm$ 0.10 (A), 0.80 $\pm$ 0.13 (P)\\
     This study & 3D B-mode and Power-Doppler (PD) & 400 & 3D CNN with Fusion & 0.849 $\pm$ 0.064 \\
    \hline
    \end{tabular}
\end{table}

Although the proposed method of combining B-mode and power Doppler images improves accuracy of placenta segmentation, there are a few limitations of the study which we want to highlight. The model had difficulty in making predictions along the placenta boundary, that we believe to be due to similarities in echo-texture compared to surrounding tissues and organs. The dataset in this study was from a single hospital but obtained by multiple sonographers with different machine settings. Future validation is required across multiple geographic locations, different vendors' ultrasound machines, diverse machine settings, and sonographers with diverse experience. 

Computer-aided placental segmentation is a useful step towards an in-utero pregnancy screening tool, highlighting placental structural, morphological, and volumetric features that may differ with pathology. The maps from segmentation algorithms could highlight areas of interest in 3D ultrasound by means of weighted attention, potentially providing information with clinical utility.

\section*{Conclusion}\label{Conclusions}
We proposed a multi-modal approach for fully-automated segmentation of the placenta with 3D ultrasound images combining B-mode and power Doppler ultrasound volumes. Combination of these modalities allows the networks to learn diverse complementary information that improves placental segmentation accuracy. We believe this to be a valuable step towards routine automated placental screening, augmenting currently available clinical information.


\backmatter
\bmhead{Acknowledgments}\label{Ack}
The study includes computations using the computational cluster \emph{Katana} supported by the Research Technology Services at UNSW Sydney. This research is supported by grant provided by the School of Computer Science and Engineering, Faculty of Engineering, UNSW, Sydney. The authors would like to express their gratitude to sonographers at the Nepean Hospital for annotating studies.

\section*{Statements and Declarations}
\subsection*{Competing Interests}
The authors declare that they have no conflict of interest.

\subsection*{Data availability statement}
The data that support the findings of this study are not publicly available due to ethics restrictions.
\bibliographystyle{unsrtnat}
\bibliography{sn-bibliography}


\end{document}